\documentclass[preprint,showpacs,preprintnumbers,amsmath,amssymb,superscriptaddress,CJK]{revtex4}

 \usepackage{amsmath}
 \usepackage{graphicx}
 \usepackage{epstopdf}
\usepackage{epsfig}
\begin{document}
\title{An perturbation-iteration method for multi-peak solitons in nonlocal nonlinear media}
\author{Weiyi Hong}
\affiliation{Guangdong Provincial Key Laboratory of Nanophotonic Functional Materials and Devices, South
China Normal University, Guangzhou 510631, P. R. China}

\author{Bo Tian}
\affiliation{Guangdong Provincial Key Laboratory of Nanophotonic Functional Materials and Devices, South
China Normal University, Guangzhou 510631, P. R. China}

\author{Rui Li}
\affiliation{Guangdong Provincial Key Laboratory of Nanophotonic Functional Materials and Devices, South
China Normal University, Guangzhou 510631, P. R. China}

\author{Qi Guo} \email[Corresponding author's email address:
]{guoq@scnu.edu.cn}
\affiliation{Guangdong Provincial Key Laboratory of Nanophotonic Functional Materials and Devices, South
China Normal University, Guangzhou 510631, P. R. China}

\author{Wei Hu}
\affiliation{Guangdong Provincial Key Laboratory of Nanophotonic Functional Materials and Devices, South
China Normal University, Guangzhou 510631, P. R. China}
\begin{abstract}
An perturbation-iteration method is developed for the computation of the Hermite-Gaussian-like solitons with arbitrary peak numbers in nonlocal nonlinear media. This method is based on the perturbed model of the Schr\"{o}dinger equation for the harmonic oscillator, in which the minimum perturbation is obtained by the iteration. This method takes a few tens of iteration loops to achieve enough high accuracy, and the initial condition is fixed to the Hermite-Gaussian function. The method we developed might also be extended to the numerical integration of the Schr\"{o}dinger equations in any type of potentials.


\end{abstract}

\maketitle

\section{Introduction}

Nonlocal spatial optical solitons have been a hot topic in the current research on nonlinear optics in the past few decades. In the Snyder and Mitchell's work\cite{A. W. Snyder-Science-97}, the nonlocal nonlinear Schr\"{o}dinger equation (NNLSE), which models the optical beam propagation in the nonlocal nonlinear media\cite{Assanto-book,Qi Guo-book-01}, can be simplified to a linear equation, so-called Snyder-Mitchell (SM) model, in the strongly nonlocal case, and then an exact Gaussian-shaped stationary solution called an accessible soliton has been obtained. This pioneer work raised much attention, and many theoretical
\cite{A. W. Snyder-Science-97,Assanto-book,Qi Guo-book-01,A.W. Snyder-josab-97,A.W. Snyder-ijstqe-00,C. Conti-prl-03} and experimental\cite{Assanto-book,Qi Guo-book-01,C. Conti-prl-04,M. Peccianti-ol-02,M. Peccianti-apl-02,M. Peccianti-pre-03} works have been focused on the propagation of nonlocal spatial solitons.

It is well known that the complex-form solitons with multi-peak structure cannot be self-guided in the local nonlinear media due to the natural repulsion existing between lobes of opposite phase. It has been found by Christodoulides et al. in 1998 that multi-peak solitons are possible in saturable nonlinear media\cite{Christod-prl-98}. Meanwhile, it is also found in nonlocal nonlinear media that the repulsion can be overcome by the nonlocality\cite{A. V. Mamaev-pra-97,X. Hutsebaut-oc-04,A. I. Yakimenko-pre-06}, and the multi-peak solitons have been observed both numerically \cite{D. W. McLaughlin-pd-96,D. Deng-josab-07,D. Buccoliero-prl-07}and experimentally\cite{X. Hutsebaut-oc-04,C. Rotschild-ol-06}, and their stability has also been discussed\cite{Z. Xu-ol-05}. In the strongly nonlocal case, Multi-peak solitons can also be analytically obtained based on SM model\cite{D. Deng-josab-07,Dongmei-jpb-2008,Dongmei-ol-2007,Dongmei-ol-2009}.

Generally, most of the solitons defy analytical expressions and have to be computed numerically. So far, a number of numerical methods have been developed. Examples include the shooting method\cite{J. Yang-pre-02}, the Petviashvili-type method\cite{V.I. Petviashvili-pp-76,M.J. Ablowitz-ol-05,T.I. Lakoba-jcp-07}, the imaginary time evolution method and its accelerated version\cite{J.J. Garcia-Ripoll-sjsc-01,J. Yang-sam-08}, the Newton's method\cite{J.P. Boyd-book-01,J.P. Boyd-jcp-02,William H. Press-book-92}, the squared-operator iteration method and its modified version\cite{J. Yang-sam-07}, Newton-conjugate-gradient (Newton-CG) methods\cite{J. Yang-jcp-09}, etc. Generally, the first three methods can only converge to the ground states\cite{T.I. Lakoba-jcp-07,J. Yang-sam-08,D.E. Pelinovsky-sjna-04,J. Yang-prl-99}, i.e., the single-peak solitons. The Newton's method has been widely used for iterating both single- and multi-peak solitons. However, it requires that the initial condition should be close enough to the exact solution\cite{William H. Press-book-92}. The squared-operator iteration method and its modified version, based on the idea of time-evolving a ``squared'' operator equation, can converge to any solitons, including multi-peak solitons\cite{J. Yang-sam-07}. However, these methods are quite slow, especially when the propagation gets near the edge of the continuous spectrum (detailed discussions can be found in Ref.\cite{J. Yang-jcp-09} ). The Newton-CG method is based on Newton iteration, coupled with conjugate-gradient iterations to solve the resulting linear Newton-correction equation. It can be applied to compute both single- and multi-peak solitons and is faster than the other leading methods as concluded by the author. Although this method can tolerate wider ranges of initial conditions than that of the Newton method, it still require that the initial condition is reasonably close to the exact solution\cite{J. Yang-jcp-09}.

In this paper, we develop a method for the computation of the Hermite-Gauss-like solitons with arbitrary peak numbers for arbitrary degree of nonlocality (as long as the nonlocality can support the soliton). The idea is inspired by the work of Ouyang et. al.\cite{Shigen Ouyang-pre-06}, where the NNlSE is approximated to a linear equation of a perturbed harmonic oscillator, and the approximate analytical soliton solutions with the peak numbers up to three have been obtained in the second-order approximation using the standard procedure of the perturbation theory in quantum mechanics\cite{W.Greiner-book-01}. In the method we developed, named as perturbation-iteration (PI) method, we start from the perturbed model of the harmonic oscillator, determine the ``minimum'' perturbation by means of the weighted least-squares (WLS) method\cite{William H. Press-book-92}, use the formal expression of infinite-order perturbation expansions\cite{A.Messiah-book-1965} to numerically calculate the eigenfunctions and eigenvalues of the perturbed model, and iterate this perturbation to obtained the multi-peak solitons with enough high accuracy. The initial condition of this method is fixed to the Hermite-Gaussian functions, since it iterates the perturbation rather than iterating the profile directly.

\section{Methods}

We start from the (1+1)-D dimensionless NNLSE which reads\cite{Qi Guo-book-01}:
\begin{equation}\label{nnlse}
i\frac{\partial u}{\partial z}+\frac{1}{2}\frac{\partial^{2} u}{\partial x^{2}}+u\int_{-\infty} ^{\infty}R(x-\xi)|u(\xi,z)|^{2}d\xi=0,\ \ \
\end{equation}
with $u(x,z)$ the complex amplitude envelop of the optical beam, $x$ and $z$ the transverse and longitude coordinates respectively, and $R(x)$ the nonlocal response function of the media. One can look for the multi-peak-soliton solutions of Eq.~(1) with the form
\begin{equation}\label{sl}
u_{n}(x,z)=\sqrt{A_{n}}\psi_{n}(x)\exp(-i\beta_{n}z)\ (n=0,1,2,3...),
\end{equation}
where $A_{n}$ is a coefficient related to the power of soliton,  $\psi_{n}(x)$ and $\beta_{n}$ are respectively the transverse distribution and propagation constant of soliton, and $n$ is the order of soliton. Note that $A_{n}$, $\psi_{n}(x)$ and $\beta_{n}$ are all real. $n=0$ corresponds to the first-order (fundamental) soliton with one peak, $n=1$ corresponds to the second-order soliton with two peaks, and so on. Then Eq.~(\ref{nnlse}) becomes a stationary equation which reads:
\begin{equation}\label{s-nnlse}
\left[-\frac{1}{2}\frac{d^{2}}{d x^{2}}-A_{n}\int_{-\infty} ^{\infty}R(x-\xi)\psi_{n}^{2}(\xi)d\xi\right]\psi_{n}(x)=\beta_{n}\psi_{n}(x).
\end{equation}

It has been discussed in Ref. \cite{Shigen Ouyang-pre-06} that Eq.~(\ref{s-nnlse}) can be treated as a perturbed model of the harmonic oscillator, and the approximate analytical soliton solutions up to $n=2$ has been obtained there by means of the perturbation theory in Quantum Mechanics\cite{W.Greiner-book-01}. In order to apply the perturbation theory for searching the numerical soliton solution $\psi_{n}(x)$ with arbitrary order $n$, we start from a perturbed model of the harmonic oscillator with the following form
\begin{equation}\label{pho}
\left[-\frac{1}{2}\frac{d^{2}}{d x^{2}}+\frac{1}{2\mu^{4}}x^{2}+f_{n}(x)\right]\psi_{n}(x)=E_{n}\psi_{n}(x),
\end{equation}
where $\psi_{n}(x)$ and $E_{n}$ are respectively the eigenfunctions and the eigenvalues of Eq.~(\ref{pho}), $f_{n}(x)$ is the perturbation compared to the term $x^{2}/(2\mu^{4})$ (the potential of the harmonic oscillator), and is expressed as $f_{n}(x)=-A_{n}\int_{-\infty} ^{\infty}R(x-\xi)\psi_{n}^{2}(\xi)d\xi-V_{n0}-x^{2}/(2\mu^{4})$ with $V_{n0}$ a constant. The eigenfunction series and the eigenvalues of the unperturbed equation for the harmonic oscillator, which is
\begin{equation}\label{UEHO}
\left(-\frac{1}{2}\frac{d^{2}}{d x^{2}}+\frac{1}{2\mu^{4}}x^{2}\right)\phi_{n}^{(0)}=\varepsilon_{n}^{(0)}\phi_{n}^{(0)},
\end{equation}
are~\cite{W.Greiner-book-01} the Hermite-Gaussian function (HGF) $\phi_{n}^{(0)}(x)=N_{n}H_{n}(x/\mu)\exp(-x^{2}/2\mu^{2})$ and $\varepsilon_{n}^{(0)}=(n+1/2)/\mu^{2}$ respectively, where $H_{n}(x)$ are the $n$-order Hermite polynomials and $N_{n}=(\mu\sqrt{\pi}2^{n}n!)^{-1/2}$ are normalization coefficients such that $\int^{\infty}_{-\infty}[\phi_{n}^{(0)}(x)]^{2}dx=1$. We emphasize here the meaning of the function $f_n(x)$: mathematically, $f_n(x)\psi_{n}(x)$ is the difference between Eq.~(\ref{s-nnlse}) and Eq.~(\ref{UEHO}) at the zero-order approximation about the perturbation such that $\psi_{n}(x)\approx \phi_{n}^{(0)}$ and $V_{n0}=\beta_n-\varepsilon_{n}^{(0)}$; physically, $f_n(x)$ is the difference between the soliton-induced nonlinear refraction $A_{n}\int_{-\infty} ^{\infty}R(x-\xi)\psi_{n}^{2}(\xi)d\xi$ and the parabolic refraction $-V_{n0}-x^{2}/(2\mu^{4})$.

For the given $f_{n}(x)$, $E_{n}$ and $\psi_{n}$ can be obtained using the standard procedure of the perturbation theory\cite{W.Greiner-book-01}. The formal expressions of the infinite-order perturbation expansions of $E_{n}$ and $\psi_{n}$ can be respectively expressed as\cite{A.Messiah-book-1965}:
\begin{equation}\label{eigenvalue}
E_{n}=\varepsilon_{n}^{(0)}+\sum_{m=1}^{\infty}
\varepsilon_{nn}^{(m)},
\end{equation}
\begin{equation}\label{wavefunction}
\psi_{n}(x)=\phi_{n}^{(0)}(x)+\sum_{m=1}^{\infty}
\phi_{n}^{(m)},
\end{equation}
and
\begin{equation}\label{eigenvaluemodify}
\phi_{n}^{(m)}=\sum_{k=0 (k\neq n) }^{\infty}\zeta_{mk}\phi_{k}^{(0)}(x)
\end{equation}with $\varepsilon_{kn}^{(m)}=\int_{-\infty}^{+\infty}\phi_{k}^{(0)}f_{n}\phi_{n}^{(m-1)}dx$,
$\zeta_{mk}=(\sum\limits_{s=0}^{m-1}\varepsilon_{nn}^{(m-s)}\zeta_{sk}-\varepsilon_{kn}^{(m)})/(\varepsilon_{k}^{(0)}-\varepsilon_{n}^{(0)})$ and $\zeta_{0k}=0$, where the summation indices $m$ and $k$ in Eqs.(\ref{eigenvalue})-(\ref{eigenvaluemodify}) are respectively the order of the expansion and the order of the HGF. Note that although the convergency of the expansions (\ref{eigenvalue}) and (\ref{wavefunction}) is lacking in mathematical rigor, our numerical computations (presented in the next section) indicate that the expansions are in physical precision.

\begin{figure}[htbp]
\centering
\includegraphics[width=10cm]{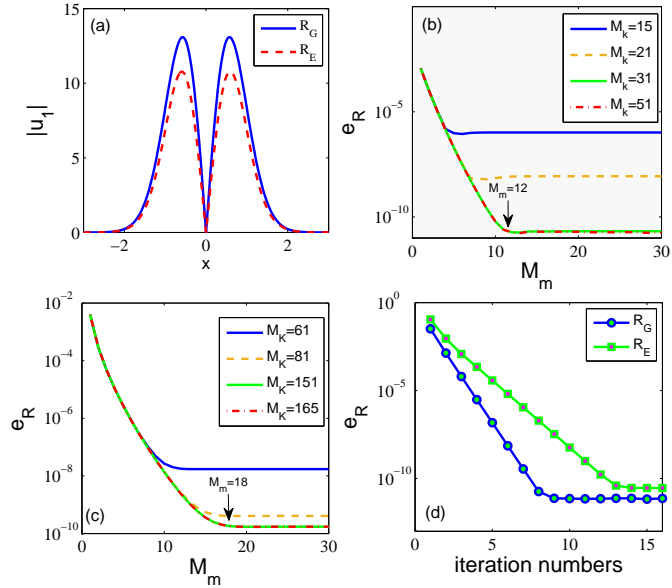}
\caption{The amplitude distributions of the solitons for the two responses (a). $e_{R}$ versus $M_{m}$ for different $M_{k}$ respectively for $R_{G}(x)$ (b) and $R_{E}(x)$ (c). The $e_{R}$ diagrams versus the iteration numbers (d).}
\end{figure}

In the following, we are going to search a $\psi_{n}(x)$ under the expected degree of the nonlocality $\sigma=w_{mat}/w_s$,
with $w_{mat}$ the characteristic length of the media and $w_s$ the root-mean-square(RMS) width of the soliton (expressed as $w_s=\sqrt{\int_{-\infty}^{+\infty}2x^{2}|\psi_{n}(x)|^{2}dx/\int_{-\infty}^{+\infty}|\psi_{n}(x)|^{2}dx}$). Since $w_s$ is unknown, we can use the RMS width of its zeroth-order approximation [$\phi_{n}^{(0)}(x)$] instead, which is $w=\mu\sqrt{2n+1}$\cite{W.Greiner-book-01}, and then give an apriori value of $\sigma$ which reads
\begin{equation}\label{sigma}
\sigma=\frac{w_{mat}}{\mu\sqrt{2n+1}}.
\end{equation}
The reason is that $w_s$ mainly depend on $\phi_{n}^{(0)}$ according to the expansion~(\ref{wavefunction}), and the higher-order corrections in the expansion only slightly deviate the RMS width $w_s$ from $w$, as will be discussed below. Therefore, by setting $\mu=w_{mat}/(\sigma\sqrt{2n+1})$, we determine a ``minimum'' $f_{n}(x)$ by means of the WLS method\cite{William H. Press-book-92}, of which the procedure is as follows. For a given transverse distribution of the optical beam $\sqrt{A_n}\psi_{n}(x)$, the deviation, which is sampled by the optical beam, between the nonlinear refraction and the parabolic refraction is
\begin{equation}\label{deltaa}
\delta(A_{n},V_{n0})=\int_{-\infty}^{\infty}f_{n}^{2}(x)\psi_{n}^{2}(x)dx.
\end{equation}
By letting the first derivative of $\delta(A_{n},V_{n0})$ with respect to $A_{n}$ and that to $V_{n0}$ equal zero respectively, we have
 \begin{equation}\label{set}
\begin{cases}
aA_n-bV_{n0}=d\\
bA_n-cV_{n0}=e,
\end{cases}
\end{equation}
where $a=\int_{-\infty}^{\infty}V^{2}_n(x)\psi_{n}^{2}(x)dx$, $b=\int_{-\infty}^{\infty}V_n(x)\psi_{n}^{2}(x)dx$, $c=\int_{-\infty}^{\infty}\psi_{n}^{2}(x)dx$, $d=\int_{-\infty}^{\infty}x^{2}V_n(x)\psi_{n}^{2}(x)dx /(2\mu^{4})$, $e=\int_{-\infty}^{\infty}x^{2}\psi_{n}^{2}(x)dx/(2\mu^{4})$, and $V_n(x)=-\int_{-\infty} ^{\infty}R(x-\xi)\psi_{n}^{2}(\xi)d\xi$.
By easily solving this equation set, $A_{n}$ and $V_{n0}$ can be obtained as
\begin{equation}\label{AV}
A_{n}=\frac{dc-be}{ac-b^{2}},~~
V_{n0}=\frac{bd-ae}{ac-b^{2}}.
\end{equation} Then, the perturbation $f_{n}(x)$ which makes the deviation $\delta$ reaches its minimum is obtained through the relationships of Eq.~(\ref{AV}). As a result, we can calculate
$E_n$ and $\psi_n$ via Eqs.~(\ref{eigenvalue}) and (\ref{wavefunction}) by truncating the summations by the maxima of $m$ and $k$ (marked as $M_{m}$ and $M_{k}$, respectively) when the error of the numerical solution can hardly decreases.

\begin{figure}[htbp]
\centering
\includegraphics[width=9.5cm]{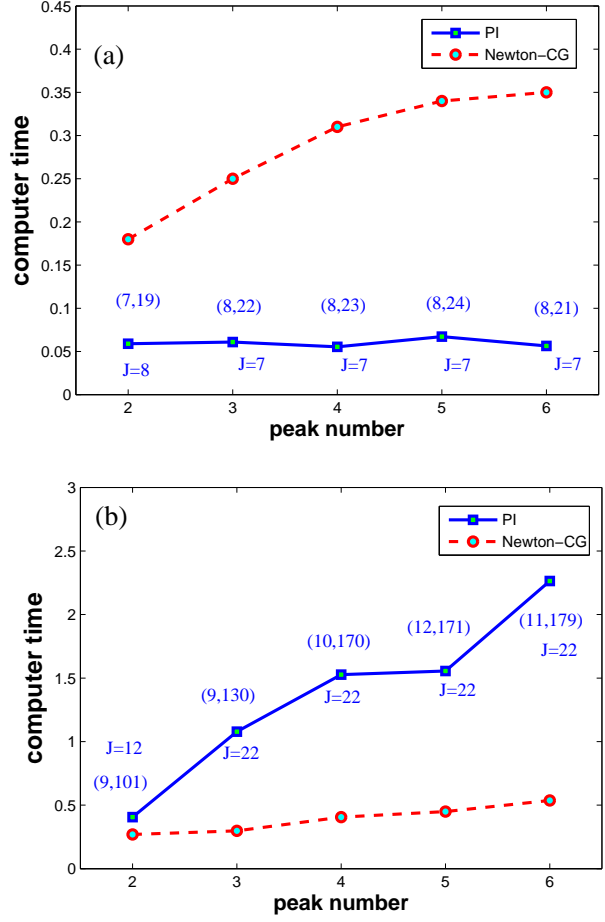}
\caption{The comparisons of the computer times of the solitons with different $n$, taken by our method (PI) and the Newton-CG method, for $R_{G}(x)$ (a) and $R_{E}(x)$ (b), respectively. The needed values of the pair ($M_m$, $M_k$) and the iteration number $J$ for every soliton are marked.}
\end{figure}

We can repeat the above procedure to achieve higher accuracy by the iteration. First, we let $\psi_{n}^{(1)}=\phi_{n}^{(0)}$ as an input wave function [$\psi_{n}^{(j)}$ is the real wavefunction in the $j$-$th$ loop$(j=1,2,3...)]$, and algorithmize this procedure according to Eqs.~(\ref{eigenvalue}), (\ref{wavefunction}), 
and (\ref{AV}) through the following iterating process: within the $j$-$th$ loop, we have
\begin{equation}\label{iteration-process-1}
A_{n}^{(j)}=\frac{d^{(j)}c^{(j)}-b^{(j)}e^{(j)}}{a^{(j)}c^{(j)}-(b^{(j)})^{2}},
\end{equation}
\begin{equation}\label{iteration-process-11}
V_{n0}^{(j)}=\frac{b^{(j)}d^{(j)}-a^{(j)}e^{(j)}}{a^{(j)}c^{(j)}-(b^{(j)})^{2}},
\end{equation}
\begin{equation}\label{iteration-process-2}
f_{n}^{(j)}=A_{n}^{(j)}V_{n}^{(j)}-V_{n0}^{(j)}-x^{2}/(2\mu^{4}),
\end{equation}
\begin{equation}\label{iteration-process-3}
E_{n}^{(j+1)}=\varepsilon_{n}^{(0)}+\sum_{m=1}^{M_{m}}\int_{-\infty}^{+\infty}\phi_{n}^{(0)}f_{n}^{(j)}\phi_{n}^{(m-1)}dx,
\end{equation}
\begin{equation}\label{iteration-process-4}
\psi_{n}^{(j+1)}=\phi_{n}^{(0)}+\sum_{m=1}^{M_{m}}\sum_{k=0 (k\neq n) }^{M_{k}}\zeta_{mk}^{(j)}\phi_{k}^{(0)}.
\end{equation}
 At last, the soliton solutions of Eq.~(\ref{nnlse}) are obtained as $u_n(x,z)=\sqrt{A_{n}^{(J)}}\psi_{n}^{(J+1)}(x)\exp(-i\beta_{n}z)\ $ with $\beta_{n}=E_{n}^{(J+1)}+V_{n0}^{(J)}$, where $J$ is the number of iteration. Note that $E_n^{j+1}$ inside the loop is used to compute the error within the loop. In principle, $M_k$ for every $m$-order of the expansion(\ref{iteration-process-4}) can be individually chosen. For the practical convenience, however, $M_k$ is set to the same value for every $m$-order in our computation below.
.

For the obtained soliton solution with the form given by Eq.~(\ref{sl}), the corresponding residual is
\begin{equation}\label{error}
\begin{aligned}
e_{A}&=\left|\frac{\sqrt{A_{n}}}{2}\frac{d^{2}}{dx^{2}}\psi_{n}(x)-A_{n}^{3/2}\psi_{n}(x)V_n(x)+\sqrt{A_{n}}\beta_{n}\psi_{n}(x)\right|\\
&=\sqrt{A_{n}}e_{R},
\end{aligned}
\end{equation}
with
\begin{equation}\label{error2}
e_{R}=\left|\frac{1}{2}\frac{d^{2}}{dx^{2}}\psi_{n}(x)-A_{n}\psi_{n}(x)V_n(x)+\beta_{n}\psi_{n}(x)\right|.
\end{equation}
 Although the residual is the function of $x$, we consider its average value within the region quadruple larger than the width of the soliton, since the data in the domain far from the soliton are meaningless for discussing the error. Obviously, $e_{A}$ consists of $\sqrt{A_{n}}$ and $e_{R}$ that we defined as the relative residual, and $e_{R}$ is more reasonable than $e_{A}$ to be used to discuss the error of the numerical results, since $\sqrt{A_{n}}$ of the solitons with different $n$ are quite different. The iteration process are terminated when the criterion $\lambda \left(\lambda=\sqrt{\int_{-\infty}^{\infty}[\psi_{n}^{(j+1)}(x)-\psi_{n}^{(j)}(x)]^{2}dx} \right)$ is sufficiently small.

\section{Examples}

\begin{figure}[htbp]
\centering
\includegraphics[width=9cm]{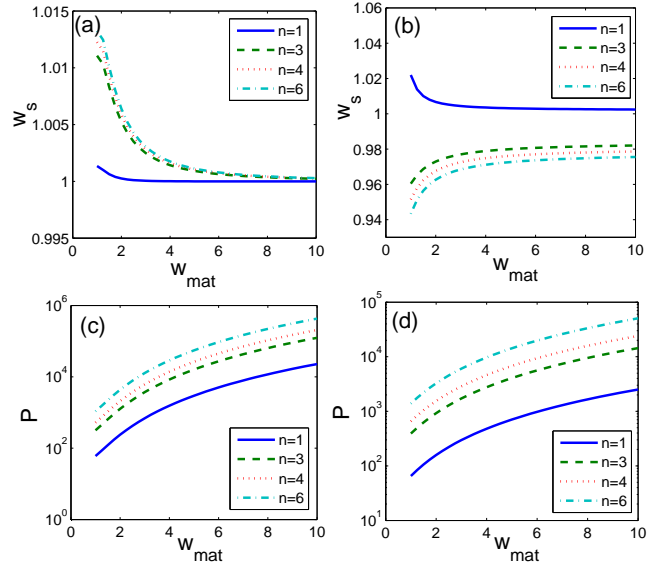}
\caption{The widths of the solitons with different $n$ as the function of $w_{mat}$ for $R_{G}(x)$ (a) and $R_{E}(x)$ (b), respectively. The power of the solitons(shown in logarithmic coordinate) with different $n$ as the function of $w_{m}$ for $R_{G}(x)$ (c) and $R_{E}(x)$ (d), respectively.}\label{Fig3}.
\end{figure}

\begin{figure}[htbp]
\centering
\includegraphics[width=9cm]{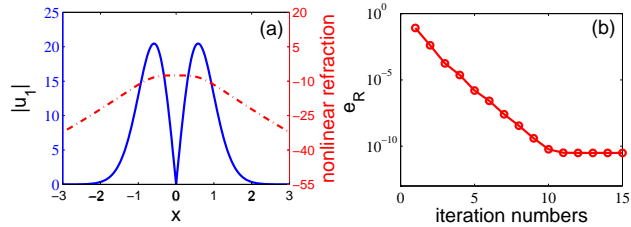}
\caption{(a) The amplitude distribution of the 2-peak soliton (blue solid curve) and the distribution of its induced nonlinear refraction (red dashed curve) of the sine-oscillation response. (b) $e_{R}$ versus the number of iteration. }\label{Fig4}.
\end{figure}

We show in Fig.~1 the computation of $2$-peak solitons ($n=1$) for Gaussian and exponential-decay response functions which respectively reads\cite{Qi Guo-book-01,Qi Guo-book-02} $R_{G}(x)=\exp[-x^{2}/(2w_{mat}^{2})]/(\sqrt{2\pi}w_{mat})$ and $R_{E}(x)=\exp(-|x|/w_{mat})/(2w_{mat})$, where $w_{mat}$ is the characteristic length of the media. In our computation, for convenience, we let $\mu=1/\sqrt{2n+1}$ to make the RMS widths of $\phi_{n}^{(0)}$ equal to $1$, and then $\sigma=w_{mat}$ [see Eq.~(\ref{sigma}) above)], which is chosen to be $2$ in this example. The computational domain is taken as $-20<x<20$, discretized by 8192($2^{13}$) points (The below computations are the same with these parameters). Figure~1(a) are the amplitude distributions of the solitons for the two responses. In order to chose proper $M_m$ and $M_k$, we present the $e_{R}$ diagrams versus $M_m$ for different $M_k$ in Figs.~1(b) and (c) respectively for $R_{G}(x)$ and $R_{E}(x)$. Here we terminate the iterations when $\lambda$ is sufficiently small. For both two cases, it is clear that $e_{R}$ rapidly declines when $M_m$ increases and becomes nearly unchangeable. As $M_k$ keeps on increasing, $e_{R}$ also declines and becomes unchangeable. The results above indicate the best accuracy can be achieved when $M_m$ and $M_k$ are both large enough. In detail, for the $2$-peak solitons, $M_m=12$ and $M_k=31$ are found for $R_G(x)$ to reach the best accuracy of $2.1\times 10^{-11}$ and $M_m=18$ and $M_k=151$ for $R_E(x)$ to reach $1.5\times 10^{-10}$. However, for other achievable accuracies one required, it is difficult to obtain the proper values of $M_m$ and $M_k$ in a fast way. One can start for small values of $M_m$ and $M_k$ (typically, $M_m=3$ and $M_k=15$), fix the value of $M_m$ and increase $M_k$, until $e_R$ reaches the required value;  if $e_R$ cannot reach the required value, then increase $M_m$ and repeat the procedure above. By setting $M_m$ and $M_k$ for the best accuracy mentioned above, we further calculate the $e_{R}$ diagrams versus the number of iterations shown in Fig.~1(d) for the cases of the two responses. It is shown that the $e_{R}$ rapidly declines with the iteration, and becomes nearly unchangeable when the number of the iteration is larger than $9$ and $13$ respectively for the cases of $R_{G}(x)$ and $R_{E}(x)$.

Our method can be used to calculate multi-peak solitons with an arbitrary order of soliton $n$. For the given $e_{R}=10^{-9}$, the computer times for the solitons with different $n$ are shown in Figs.~2(a) and (b) for $R_{G}(x)$ and $R_{E}(x)$, respectively. In addition, the needed values of the pair ($M_m$, $M_k$) and the iteration number $J$ for every soliton are marked in the figures. For comparison, the results taken by Newton-CG method\cite{J. Yang-jcp-09} are also presented. It is shown that our method takes even shorter times than the Newton-CG method for $R_G(x)$, but much longer times for $R_E(x)$, since the case of $R_E(x)$ takes much larger values of $M_m$, $M_k$ and $J$ than that of $R_G(x)$. Even so, our method is quite accessible, since the cost computer times are in the order of seconds, and the initial condition is fixed to $\phi_{n}^{(0)}(x)$, rather than careful choosing of an initial profile. Note that the Newton-CG method requires that the initial condition is reasonably close to the exact solution\cite{J. Yang-jcp-09}. In our realization of the Newton-CG method, therefore, we choose $B\phi_{n}^{(0)}(x)$ (with $B$ the initial amplitude) as the initial condition, and $B$ should be carefully chose to ensure the convergence of the iteration. Moreover, the sensitivity of $B$ also increases as the peak number increases, according to our computation.

We also present the RMS widths of the solitons $w_s$ with different $n$ as a function of the characteristic length $w_{mat}$, i.e, the degree of the nonlocality, for $R_{G}(x)$ and $R_{E}(x)$, which are shown in Figs.~3(a) and~(b), respectively. The request values of $e_{R}$ are the same with those in Fig.~2. As expected above, the higher-order corrections in the expansion~(\ref{wavefunction}) only slightly deviate the RMS widths from $1$. As $w_{mat}$ increases, $w_s$ for the case of $R_{G}(x)$ tends to $1$, while that for the case of $R_{E}(x)$ tends to a value deviated from $1$. This asymptotic behavior can be understood as follows\cite{Qi Guo-book-01,Qi Guo-book-02}: as $w_{mat}$ increases, namely, the degree of the nonlocality increases, Eq.~(\ref{s-nnlse}) with $R_{G}(x)$ tends to the SM model, i.e., the model for harmonic oscillator described by Eq.~(\ref{UEHO})~\cite{A. W. Snyder-Science-97}, and then $w_s$ tends to $1$; for the case of $R_{E}(x)$, in the strongly nonlocal limit, i.e., $w_{mat}\rightarrow \infty$, Eq.~(\ref{s-nnlse}) does not tend to Eq.~(\ref{UEHO}) due to the discontinuity in the derivative of $R_{E}(x)$ at the origin (detailed discussions can be found in \cite{Qi Guo-book-01,Qi Guo-book-02}), therefore this deviation is in expectation. It is worth mentioning that the best of achievable $e_{R}$ is not sensitive to $w_{mat}$, i.e., the degree of the nonlocality, according to our computation. Figs.~3(c) and~(d) are the power of the solitons (shown in logarithmic coordinate) with different $n$ as the function of $w_{mat}$ for $R_{G}(x)$ and $R_{E}(x)$, respectively. As $n$ and $w_{mat}$ increase, the power of the solitons increases rapidly, which is the reason why $e_{R}$ is more proper than $e_{A}$ to discuss the error of the solitons, as mentioned above.

Finally, we show in Fig.~4 the computation of the 2-peak soliton with the sine-oscillation response $R_{S}(x)=-\sin(|x|/w_{m})/(2w_{m})$\cite{sine1,sine2}, which could hardly be obtained by other known methods as far as we have tried. The best accuracy of $3.2\times10^{-11}$ is achieved when $M_m=25$ and $M_k=179$. The physical properties for such multi-peak solitons with $R_{S}(x)$ will be further discussed otherwhere.

\section{Conclusion}

In this work, we develop a method for the numerical computations of the Hermite-Gauss-like solitons with arbitrary peak numbers in nonlocal nonlinear media. We start from the perturbed model of the harmonic oscillator, determine the ``minimum'' perturbation by means of the WLS method, and use the formal expression of infinite-order perturbation expansions to numerically calculate the eigenfunctions and eigenvalues of the perturbed model. Finally we iterate the procedure above to achieve a sufficient high accuracy. This method takes a few tens of iteration loops, and the initial condition is fixed to the Hermite-Gaussian function. It is worth mentioning that our method is not restricted to the NNLSE, but might also be able to be extended to the Schr\"{o}dinger equations in any type of potentials.

\section{Funding Information}

National Natural Science Foundation of China (Grant No. 11474109).


\begin{thebibliography}{200}
\bibitem{CG2}
In the Newton-CG method, the residuals diverges for the sine oscillation response as far as we can try.
\end{thebibliography}
\end{document}